\documentclass[10pt,aps,pra,twocolumn]{revtex4-1}
\usepackage{graphicx}
\usepackage{ulem}
\usepackage{amsmath}
\usepackage{amssymb}
\usepackage{wrapfig}
\usepackage[utf8]{inputenc}  
\usepackage{color} 

\begin{document}

\title{Modulation spectroscopy of Rydberg atoms in an optical lattice}
\author{V. S. Malinovsky$^{1,2}$}
\author{K.~R.~Moore$^{3, \dag}$}
\author{G.~Raithel$^{3}$}
\thanks{Corresponding author: graithel@umich.edu}
\affiliation{$^1$US Army Research Laboratory, Adelphi, MD 20783}
\affiliation{$^2$Department of Physics, Stevens Institute of Technology, Hoboken, New Jersey 07030}
\affiliation{$^3$Department of Physics, University of Michigan, Ann Arbor, Michigan 48105}
\date{\today}

\begin{abstract}
We develop and study quantum and semi-classical models of Rydberg-atom spectroscopy in amplitude-modulated optical lattices. Both initial- and target-state Rydberg atoms are trapped in the lattice. Unlike in any other spectroscopic scheme, the modulation-induced ponderomotive coupling between the Rydberg states is spatially periodic and perfectly phase-locked to the lattice trapping potentials. This leads to a novel type of sub-Doppler mechanism, which we explain in detail. In our exact quantum model, we solve the time-dependent Schr\"odinger equation in the product space of center-of-mass (COM) momentum states and the internal-state space. We also develop a perturbative model based on the band structure in the lattice and Fermi's golden rule, as well as a semi-classical trajectory model in which the COM is treated classically and the internal-state dynamics quantum-mechanically. In all models we obtain the spectrum of the target Rydberg-state population versus the lattice modulation frequency, averaged over the initial thermal COM momentum distribution of the atoms. We investigate the quantum-classical correspondence of the problem in several parameter regimes and exhibit spectral features that arise from vibrational COM coherences and rotary-echo effects. Applications in Rydberg-atom spectroscopy are discussed.
\end{abstract}

\maketitle

\section{Introduction}
\label{sec:introduction}
The interaction of an electron with an electromagnetic field consists of a term $\hat{e \bf{A}} \cdot \hat{\bf{p}} / m$ and a term $e^2 \hat{A}^2/(2m)$, with electron mass $m$, elementary charge $e$, the field's vector potential ${\bf{A}}(\hat{\bf{r}})$, and electron position and momentum operators $\hat{\bf{r}}$ and $\hat{\bf{p}}$~\cite{Friedrichbook, Sakuraibook}. Under certain conditions, periodically modulated and inhomogeneous fields can drive electronic transitions via a ponderomotive interaction, $e^2 \hat{A^2}/(2m)$.
These transitions can occur in Rydberg atoms immersed in light fields, a case in which the field frequency (hundreds of THz) exceeds the Rydberg atom's evolution frequency (tens to hundreds of GHz) by several orders on magnitude.
Due to the quasi-free nature of the Rydberg electron on optical time- and energy-scales, the Rydberg-atom ponderomotive effect is related to free-electron Kapitza-Dirac scattering~\cite{Batelaan2000, Kozak2018, Freimund2001}. Further, ponderomotive level shifts in high-intensity laser fields were observed in atoms~\cite{Normand1989, OBrian1994} and in molecules~\cite{LopezMartens2000}, as well as in high-intensity multi-photon ionization \cite{Helm1991}, in high-intensity zero-kinetic-energy photoelectron (ZEKE) spectroscopy~\cite{Zavriyev1995}, and in optical~\cite{Paulus2001,Kopold2002} and microwave~\cite{Gallagher1989,
Arakelyan2016} above-threshold ionization. Ponderomotive forces on atomic electrons are important in
atom dynamics in high-intensity laser pulses~\cite{Eichmann2009, Chen2017, Cai2019}.
Ponderomotive effects are also known from Paul ion traps, where the ponderomotive force drives the secular motion, while the micromotion occurs at the trap's radio-frequency drive~\cite{Bollinger1994}.

In Rydberg atoms, ponderomotively driven transitions (transitions driven by the $e^2 \hat{A^2}/(2m)$ operator) are free of multipole selection rules~\cite{Knuffman2007} that govern traditional methods of laser and microwave spectroscopy (which are based on the properties of the $\hat{e \bf{A}} \cdot \hat{\bf{p}} / m$ operator in first or higher orders). For ponderomotive spectroscopy to be effective, the field must be modulated at a (sub-)harmonic of the transition frequency, and the field intensity must vary within the extent of the electron wavefunction. These conditions are quite naturally satisfied by Rydberg-atom transitions in amplitude- or phase-modulated optical lattices~\cite{Knuffman2007}, because Rydberg atoms have sizes on the order of typical optical-lattice periods~\cite{GallagherBook}. Further, microwave amplitude and phase modulators for optical-lattice lasers are readily available.
In addition to driving microwave transitions, the ponderomotive interaction can serve as a tool to trap the Rydberg atoms in a ponderomotive optical lattice (POL)~\cite{Dutta2000,Younge2010,YoungeNJP,Anderson2011,Topcu2013}.
Hence, the modulated POL can satisfy two functionalities at once: it can trap the Rydberg atoms and, at the same time, serve as a spectroscopic probe for a wide variety of Rydberg transitions~\cite{Moore2015}.
POLs also offer great flexibility in designing Rydberg-state-mixing properties~\cite{Younge2010b,Anderson2012,Wang2016} and magic-transition traps, where two or more states have (near-) identical trapping potentials~\cite{MooreMagic}.
In these applications perturbing effects from Rydberg-atom photo-ionization~\cite{Saffman2005,Zhang2011,Tallant2010,Markert2011,Anderson2013} are typically irrelevant, in particular for $S$-type Rydberg states of rubidium and for high-angular-momentum Rydberg states of any species.

The combination of the aforementioned features enables high-precision spectroscopy on long-lived circular-state Rydberg atoms trapped in optical lattices~\cite{Ramos2017}, utilizing a scheme in which the transition frequency between the circular Rydberg levels is measured via resonant POL-modulation at microwave frequencies. This experimental platform may also be useful for quantum simulators~\cite{Nguyen2018} that are based on circular-state Rydberg-atom arrays.
For such applications, the effects of lattice-potential-induced level shifts and interaction-time broadening on the
achievable spectroscopic resolution have to be reduced. To that end, it is important to quantitatively model vibrational sidebands, anharmonic corrections, and vibrational quantization and tunneling in POL modulation spectroscopy. This necessitates a fully quantized description of the center-of-mass (COM) and internal-state dynamics of the atoms.
In the present paper we develop such models, investigate quantum-classical correspondence in POL modulation spectroscopy, and exhibit the quantum features in the spectra.

\section{Overview of lattice modulation spectroscopy}
\label{sec:overview}

\begin{figure}[h]
\centerline{\includegraphics[width=9.5cm]{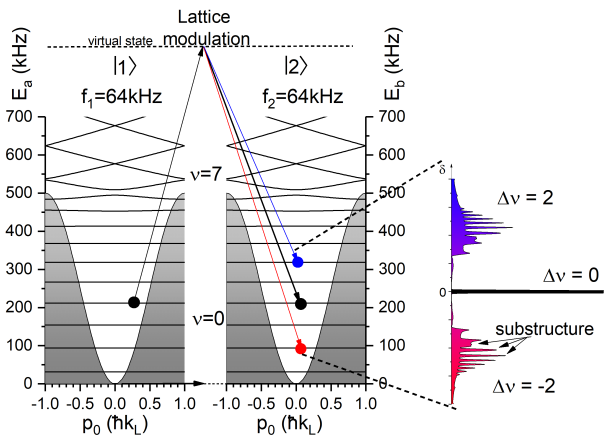}}
\caption{(Color online) Band structures and even-parity transitions between Rydberg states labeled $\vert 1 \rangle$ (left) and $\vert 2 \rangle$ (right) in identical, modulated, ponderomotive optical-lattice potentials. The spectrum consists of a Doppler-free central peak (change in vibrational quantum number $\Delta \nu =0$) and vibrational sidebands ($\Delta \nu = \pm 2$). In the depicted case, the anharmonicity-induced substructure of the sidebands is spectroscopically resolved.  }\label{fig:g1}
\end{figure}

In Fig.~\ref{fig:g1} we illustrate several quantum aspects of POL modulation spectroscopy. The curves with the light-gray drop areas visualize the sinusoidal COM lattice potentials versus position for a pair of Rydberg states $\vert 1 \rangle$ and $\vert 2 \rangle$; the figure shows the case of a magic POL transition for a lattice depth of 500~kHz. The depicted band structure of the COM dynamics is for $^{85}$Rb atoms in a POL formed from two counter-propagating beams of 1064~nm wavelength. Amplitude modulation (AM) of the lattice drives electronic transitions the Rabi frequencies of which have previously been calculated~\cite{Knuffman2007}. While POL modulation spectroscopy has no selection rules for $l^{\prime} - l$ ($l$ and $l^{\prime}$ are the angular-momentum quantum numbers of Rydberg states $\vert 1 \rangle$ and $\vert 2 \rangle$, respectively), for even-parity transitions ($|l^{\prime} - l| = 0, 2, \ldots$) the vibrational quantum number $\nu$ of the COM motion can only be changed by even numbers, whereas for odd-parity transitions ($|l^{\prime} - l| = 1, 3, \ldots$) it can only be changed by odd numbers. The case depicted in Fig.~\ref{fig:g1} is for even-parity transitions. The spectrum, sketched on the right, shows the transition probability from $\vert 1 \rangle$ into $\vert 2 \rangle$ as a function of the detuning of the POL modulation frequency, $\delta$, from the atomic transition frequency. The spectrum has red-shifted ($\Delta \nu = \nu'- \nu = -2$), unshifted ($\Delta \nu = 0$) and blue-shifted ($\Delta \nu = 2$) spectral components, which arise from transitions of the lattice-trapped atoms. The substructure of the $\Delta \nu = \pm 2$ components is a COM quantum effect that results from the anharmonicity of the POL potential and that requires sufficient spectral resolution to be observed. For odd-parity transitions,
the spectrum would have two major components, corresponding to changes of the vibrational quantum number by $\Delta \nu = \pm 1$.

The theory leading to spectra as sketched in Fig.~\ref{fig:g1} is developed and discussed in Secs.~\ref{sec:SE}-\ref{sec:sc}. In our models both ground and excited wave functions evolve on one-dimensional, sinusoidal optical-lattice trapping potentials~\cite{Dutta2000} with generally different depths but fixed relative spatial phase (the experimentally most relevant case). The ponderomotive coupling between the electronic states that arises from lattice modulation is described by an effective Rabi frequency, $\Omega(z)$, that depends on the atom's COM position $z$ transverse to the lattice planes. The novelties described in this work depend critically on the fact that $\Omega(z)$ has a sine-like dependence on $z$, with the same spatial period as the lattice itself~\cite{Knuffman2007}. The spatial phase between the ponderomotive coupling and the lattice depends on whether the modulation-driven transition is between Rydberg states with same or opposite parity. The phase of the ponderomotive coupling $\Omega(z)$ exhibits a quite peculiar behavior, as it proceeds in discrete steps of $\pi$ as a function of the coordinate $z$. This behavior differs radically from the optical phase of typical plane-wave or Raman couplings, $\Delta \bf{k} \cdot \bf{r}$, which is a continuous function of position ($\hbar \Delta \bf{k}$ is the photon momentum transfer). In Sec.~\ref{subsec:g2b} we explain why the peculiar phase behavior of $\Omega(z)$ in modulated POLs leads into a new paradigm of sub-Doppler spectroscopy.

To obtain the spectrum of the excited-state population as a function of lattice modulation frequency, we numerically solve the time-dependent Schr\"odinger equation (TDSE) in momentum representation of the COM. The results of these numerical solutions are averaged over the thermal COM momentum distribution of the Rydberg atoms. In addition, we analyze the band structure of the problem. The band structure is employed to model the POL modulation spectra with transition rates between Bloch states, averaged over an initial
thermal momentum distribution of atoms loaded into the lattice.
The solutions presented in this work account for quantum features such as band structure, vibrational quantization, band curvature and tunneling, and rotary-echo effects. In addition, we discuss the convergence of quantum and semi-classical results in the appropriate limits.
In applications, the advanced modeling afforded by our work will enable a reduction of systematic errors caused by lattice-induced shifts of the Rydberg-atom transition frequency.

\section{Time-dependent Schr{\"o}dinger equation}
\label{sec:SE}

\subsection{Position representation}
\label{subsec:SEpos}

We consider a Rydberg atom moving in a one-dimensional POL formed by two counter-propagating laser beams of equal polarization, wavenumber $k$, and wavelength $\lambda = 2 \pi /k$. The atom is initially prepared in state $\vert 1 \rangle$. The optical lattice is amplitude-modulated at a frequency that effects ponderomotively driven transitions into Rydberg state $\vert 2 \rangle$. Lattice potentials and Rabi frequencies for this scheme have been derived in~\cite{Dutta2000, Knuffman2007}; here we recite relevant, previously-proven findings. The lattice potentials for the two Rydberg states can be written in the form $V_{i,0} + V_i \cos(2kz)$, with $i=1,2$ and constants $V_{i,0} \geq V_i$. Here, it is sufficient to consider Rydberg levels $\vert n, l, j, m_j \rangle$ without lattice-induced state mixing~\cite{Dutta2000, Knuffman2007}. Assuming azimuthal symmetry and choosing the quantization axis along the direction of the laser beams, amplitude modulation of the lattice can generally drive, in first order, any transition $\vert n, l, j, m_j \rangle$ $\leftrightarrow$ $\vert n', l', j', m_j' \rangle$ with $m'_j - m_j = \Delta m_j = 0$, and with no other applicable selection rules~\cite{Knuffman2007}. To drive the transitions efficiently, the lattice modulation frequency must be close to the atomic transition frequency~\cite{Moore2015} or a sub-harmonic~\cite{MooreMagic}. The Rabi frequencies for ponderomotive transitions driven by amplitude-modulated lattices~\cite{Knuffman2007} take the form $\Omega \cos(2kz)$ for even-parity ($|l^{\prime} - l| = 0, 2, \ldots$) and
$\Omega \sin(2kz)$ for odd-parity ($|l^{\prime} - l| = 1, 3, \ldots$ ) Rydberg transitions.

We define spinor wavefunctions $\Psi_i(z)$ via
\begin{equation}
\vert \psi \rangle = \int \Psi_1(z) \vert 1, z \rangle dz + \int \Psi_2(z) \vert 2, z \rangle dz ,
\end{equation}
where the base kets $\vert 1, z \rangle$ are in the product space of the internal (Rydberg) state space $\left\{\vert 1 \rangle, \vert 2 \rangle\right\}$ and the position space $\left\{\vert z \rangle\right\}$ of the $z$-COM degree of freedom.
We further define an effective atom-field detuning $\delta = \delta_0 +V_{2,0} - V_{1,0}$, where $V_{i,0}$
are the lattice-potential offsets defined above, and
$\delta_0$ is the lattice-free atomic transition frequency, $\omega_A$, minus the optical-lattice modulation frequency, $\omega_L$, or its relevant overtone, $ p \, \omega_L$ ($p=2,3,\ldots$).
In the most generic case, considered here, only the states $\vert 1 \rangle$ and $\vert 2 \rangle$ are close to resonance, while for all other transitions  $\Omega \ll \delta$. Hence, in a dressed-atom picture the near-resonant atom-field states
are $\vert 1, N_0 \rangle$ and $\vert 2, N_0-p \rangle$, where $N_0$ is a constant number of spectator photons/phonons in the modulator crystal, and $p$ is the modulation order that drives the transition. In a holistic picture, $p$ can be interpreted as the number of light-modulator energy quanta (photons/phonons) that are absorbed in the POL-modulation transition. For the cases studied here, we neglect natural and black-body Rydberg-atom decay.

With these definitions, the Schr{\"o}dinger equation for
even-parity ($|l^{\prime} - l| = 0, 2, \ldots$ ) Rydberg transitions is
\begin{equation}\label{Master1}
\begin{array}{l}
i\dot{\Psi}_{1}= \frac{\hat{p}^2}{2\hbar m} \Psi_{1} -(\delta/2 )\Psi_{1}+ \cos[2 k z] \left(V_{1} \Psi_{1} - \Omega \Psi_{2}\right) \, ,\\
i\dot{\Psi}_{2}= \frac{\hat{p}^2}{2\hbar m} \Psi_{2} +(\delta/2 )\Psi_{2}+ \cos[2 k z] \left(V_{2} \Psi_{2} - \Omega \Psi_{1}\right) \, ,
\end{array}
\end{equation}
where the spinor wavefunctions $\Psi_i (z)$ are now in the dressed-atom picture (rotating-frame). We write detunings and couplings in units rad/s.
For odd-parity ($|l^{\prime} - l| = 1, 3, \ldots$ ) Rydberg transitions it is
\begin{equation}\label{Master-odd}
\begin{array}{l}
i\dot{\Psi}_{1}= \frac{\hat{p}^2}{2\hbar m} \Psi_{1} -(\delta/2 )\Psi_{1}+ \cos[2 k z] V_{1} \Psi_{1} - \sin[2 k z] \Omega \Psi_{2} \, ,\\
i\dot{\Psi}_{2}= \frac{\hat{p}^2}{2\hbar m} \Psi_{2} +(\delta/2 )\Psi_{2}+ \cos[2 k z] V_{2} \Psi_{2} - \sin[2 k z]\Omega \Psi_{1} \, .
\end{array}
\end{equation}
The potential depths $V_i$ are generally different. There exist cases in which $V_1 = V_2$; these are ``magic'' lattices that lead to particularly narrow spectral lines. Magic lattices are well known from optical clocks (see, for instance,~\cite{Takamoto2005,Ludlow2006}). The same concept translates to spectroscopy in modulated POLs for Rydberg atoms. Magic transitions in modulated POLs require a suitable combination of lattice period and Rydberg levels; such transitions have already been demonstrated~\cite{MooreMagic}.
Magic lattices are particularly useful for high-precision spectroscopy because they minimize lattice-induced shifts of the transition frequency to be measured. Several examples discussed below are for magic transitions, where $V_1 = V_2$. We also consider a generic case in which the lattices for $\vert 1 \rangle$ and $\vert 2 \rangle$ have different depths, $V_1 \ne V_2$.

For even-parity transitions, the Rabi frequency and the lattice potentials all share the same spatial modulation $\propto \cos(2kz)$, while  for odd-parity transitions the lattice potentials are $\propto \cos(2kz)$ and the Rabi frequency is $\propto \sin(2kz)$. In either case, the drive term $\Omega(z)$ is real and alternates between positives and negatives, amounting to discrete phase
jumps of $\pi$ at every $ \pi / (2 k)$ step in $z$.
This uncommon behavior greatly differs from the Rabi-frequency behavior
of first- and higher-order multipole transitions
effected by the $\hat{e \bf{A}} \cdot \hat{\bf{p}} / m$ term, which (in one-dimensional cases) typically is
$\Omega_m \exp( {\rm{i}} \Delta k \, z)$, where $\hbar  \Delta k$ denotes the recoil momentum and $\Omega_m$ the Rabi frequency of the transition.
We see that in the modulated-POL case the spatial phase of the drive follows a (real-valued) staircase function and the magnitude of the coupling varies in $z$ as $\vert \Omega \cos(2kz) \vert$ or $\vert \Omega \sin(2kz) \vert$, whereas in the latter case the spatial phase is a linear function with fixed slope $\Delta k$ and the
magnitude of the coupling, $\vert \Omega_m \vert$, is fixed. In Sec.~\ref{subsec:g2b} we show that these facts enable a novel type of sub-Doppler method that is realized automatically in modulated-POL spectroscopy.

\begin{widetext}

The Hamiltonian in the above  Schr{\"o}dinger equation for the even-parity case can be conveniently written in matrix form
\begin{equation}\label{Pot-Ham1-even}
\hat{H}_{even}= \frac{\delta}{2} \left(
\begin{array}{cc}
-1 & 0 \\
 0 & 1 \end{array}
\right) + \cos[2 k \hat{z}] \left(
\begin{array}{cc}
 V_{1} & -\Omega \\
 -\Omega & V_{2} \end{array}
\right)= -\frac{\delta}{2} \hat{\sigma}_{z} + \cos[2 k \hat{z}] \left( V_{+} \hat{I} - V_{-} \hat{\sigma}_{z} - \Omega \hat{\sigma}_{x}\right)
\, ,
\end{equation}
where $V_{\pm}=(V_{2}\pm V_{1})/2$, $\hat{\sigma}_{i}$ are the Pauli operators, $\hat{I}$ is the identity operator, and $\hat{z}$ is the position operator for the $z$-component of the COM motion. Similarly, for odd-parity transitions it is
\begin{equation}
\label{Pot-Ham1-odd}
\hat{H}_{odd}= -\frac{\delta}{2} \hat{\sigma}_{z} + \cos[2 k \hat{z}] \left( V_{+} \hat{I} - V_{-} \hat{\sigma}_{z} \right)
- \sin[2 k \hat{z}] \Omega \hat{\sigma}_{x}
\, .
\end{equation}

\subsection{Momentum representation}
\label{subsec:SEmom}

Due to the periodicity of potentials and couplings, the TDSE is most conveniently solved in the momentum representation~\cite{BermanBook,Malinovsky2003}. Using
\begin{equation}\label{Momentum-pr1}
\Psi_{1,2}(p)= \frac{1}{2 \pi \hbar} \int dz e^{-i\frac{p}{\hbar} z} \Psi_{1,2}(z)   \, ,
\end{equation}

\noindent we obtain the equations for the momentum-space spinor wavefunctions
\begin{equation}\label{Master-Moment-O1}
\begin{array}{l}
i\dot{\Psi}_{1}(p)= \frac{p^2}{2\hbar m} \Psi_{1}(p) -\frac{\delta}{2}\Psi_{1}(p) +
\frac{1}{2} V_{1} \left[ \Psi_{1}(p+2\hbar k)+  \Psi_{1}(p-2\hbar k)\right]
- \frac{1}{2} \Omega \left[ \Psi_{2}(p+2\hbar k)+  \Psi_{2}(p-2\hbar k)\right] \, ,\\
i\dot{\Psi}_{2}(p)= \frac{p^2}{2\hbar m} \Psi_{2} (p) + \frac{\delta}{2} \Psi_{2} (p) +
\frac{1}{2} V_{2} \left[ \Psi_{2}(p+2\hbar k)+  \Psi_{2}(p-2\hbar k)\right]
- \frac{1}{2} \Omega \left[ \Psi_{1}(p+2\hbar k)+  \Psi_{1}(p-2\hbar k)\right]  \, ,
\end{array}
\end{equation}
for the even-parity case, and similarly for the odd-parity case
\begin{equation}\label{Master-Moment-O1odd}
\begin{array}{l}
i\dot{\Psi}_{1}(p)= \frac{p^2}{2\hbar m} \Psi_{1}(p) -\frac{\delta}{2}\Psi_{1}(p) +
\frac{1}{2} V_{1} \left[ \Psi_{1}(p+2\hbar k) +  \Psi_{1}(p-2\hbar k)\right]
+ \frac{i}{2} \Omega \left[ \Psi_{2}(p+2\hbar k) -  \Psi_{2}(p-2\hbar k)\right] \, ,\\
i\dot{\Psi}_{2}(p)= \frac{p^2}{2\hbar m} \Psi_{2} (p) + \frac{\delta}{2} \Psi_{2} (p) +
\frac{1}{2} V_{2} \left[ \Psi_{2}(p+2\hbar k) +  \Psi_{2}(p-2\hbar k)\right]
+ \frac{i}{2} \Omega \left[ \Psi_{1}(p+2\hbar k) -  \Psi_{1}(p-2\hbar k)\right]  \, .
\end{array}
\end{equation}

From these equations it is seen that in momentum representation the Hilbert space breaks up into subspaces
$\left\{\vert 1 \rangle, \vert 2 \rangle\right\}$ $\otimes$ $\left\{\vert p_0 + 2 n \hbar k\rangle \, \vert \, n=0, \pm 1, \pm 2, \ldots \right\}$, with real-valued $p_0 \in \, [ - \hbar k, \hbar k ]$ (first Brillouin zone). Subspaces with different $p_0$ do not couple to each other. Using the initial condition $\Psi_{1}(p,t=0)=\delta(p-p_0- 2 n_0  \hbar k)$ and $\Psi_{2}(p,t=0)=0$, the evolution is restricted to the subspace for $p_0$. We can then set
\begin{equation}\label{Master-Moment-init-O1}
\begin{array}{l}
\Psi_{1}(p,t)=\sum_{n=-\infty }^{\infty }a_{n}(t)\delta (p-p_0-2n\hbar k) \, ,\\
\Psi_{2}(p,t)=\sum_{n=-\infty }^{\infty }b_{n}(t)\delta (p-p_0-2n\hbar k)\, ,
\end{array}
\end{equation}
and write Eq.~(\ref{Master-Moment-O1}) in matrix form as
\begin{equation}
i\left(
\begin{array}{c}
\dot{A}(t)\\
\dot{B}(t)
\end{array}
\right)=
\left(
\begin{array}{cc}
H_{A} & H_{AB}\\
H_{BA}^* & H_{B}
\end{array}
\right)
\left(
\begin{array}{c}
A(t)\\
B(t)
\end{array}
\right)
\label{eq:matrixform}
\end{equation}
where $A(t) = \left\{ \cdots , a_{-1}(t), a_{0}(t), a_{1}(t), \cdots \right\}$, $B(t) = \left\{ \cdots , b_{-1}(t), b_{0}(t), b_{1}(t), \cdots \right\}$,
and
\begin{equation}
H_{A,B} =\left\{
\begin{array}{cccccc}
\ddots  & \ddots  & 0 &  &  &  \\
\ddots  & E_{-1}^{(A,B)}/\hbar & V_{1,2}/2  & 0 &  &  \\
0 & V_{1,2}/2 & E_{0}^{(A,B)}/\hbar & V_{1,2}/2  & 0 &  \\
& 0 & V_{1,2}/2  & E_{+1}^{(A,B)}/\hbar & \ddots  &  \\
&  & 0 & \ddots  & \ddots  &
\end{array}
\right\} \, .
\label{eq:Hmatrix-O1}
\end{equation}
For even parity we have
\begin{equation}
H_{AB} = -\frac{\Omega}{2}  \left\{
\begin{array}{cccccc}
\ddots  & \ddots  & 0 &  &  &  \\
\ddots  & 0 & 1 & 0 &  &  \\
0 & 1 & 0 & 1 & 0 &  \\
& 0 & 1 & 0& \ddots  &  \\
&  & 0 & \ddots  & \ddots  &
\end{array}
\right\} \,,
\label{eq:Hmatrix-1-O1}
\end{equation}
while for odd parity we have
\begin{equation}
H_{AB} = \frac{i \Omega}{2}  \left\{
\begin{array}{cccccc}
\ddots  & \ddots  & 0 &  &  &  \\
\ddots  & 0 & 1 & 0 &  &  \\
0 & -1 & 0 & 1 & 0 &  \\
& 0 & -1 & 0& \ddots  &  \\
&  & 0 & \ddots  & \ddots  &
\end{array}
\right\} \,.
\label{eq:Hmatrix-1-O1odd}
\end{equation}

Defining $p_{0}=\beta \hbar k$ and the two-photon recoil frequency $\omega_{k}=2\hbar k^2/m$,
the lattice-free energies in Eq.~(\ref{eq:Hmatrix-O1}) can be written as
\begin{equation}
E_{n} ^{(A,B)}=  \frac{(p_0+n 2 \hbar k)^{2}}{2m} \mp \hbar \frac{\delta}{2} =
\hbar \omega_{k}(\beta/2 +n)^2 \mp \hbar \frac{\delta}{2}\, ,
\label{eq:quasiE}
\end{equation}
where $n=0, \pm 1, \pm 2, \cdots$. For $^{85}$Rb atoms in an optical lattice formed by a pair of counter-propagating 1064-nm laser beams, the case studied below, the two-photon recoil frequency is $\omega_k = 2 \pi \times 8.300$~kHz $= 52.15$~krad/s.

From the above analysis we observe that states with momentum $p$ couple only to the neighboring states $p\pm 2\hbar k$ for both $\Psi_{1}(p)$ and $\Psi_{2}(p)$. The internal-state-conserving momentum coupling strengths are $V_1/2$ and $V_2/2$ for states $\vert 1 \rangle$ and $\vert 2 \rangle$, respectively. The POL modulation induces
internal-state-changing couplings between initial- and target-level momentum states that also differ by $2\hbar k$; those couplings have a strength determined by $\Omega$.

\end{widetext}

\subsection{Averaging over the thermal momentum distribution}
\label{subsec:Max}

For reference, we state the one-dimensional Maxwell distribution in velocity, momentum and our dimensionless scaled momentum $\beta$,
\begin{eqnarray}
f (v_0)  dv_0 & = & \sqrt{\frac{m}{2\pi k_B T} }  e^{-\frac{m v_0^2}{2 k_B T}} dv_0 \nonumber \\
f (p_0)  dp_0 & = & \frac{1}{\sqrt{2\pi m k_B T} }e^{-\frac{p_0^2}{2 m k_B T}} dp_0 \nonumber \\
f (\beta)  d\beta & = & \frac{1}{\sqrt{2\pi } } \sqrt{ \frac{\hbar \omega_k}{ 2 k_B T}}
e^{-\frac{\beta^2}{4} \frac{\hbar \omega_k}{k_B T}} d\beta
\, .
\label{momentum-Dist}
\end{eqnarray}
The scaled-momentum distribution, $f (\beta)$, can be written in terms of the width of the energy distribution in units of the thermal energy. Defining $\omega_T=k_B T/\hbar$, this is $f (\beta)  = \frac{1}{\sqrt{2\pi } } \sqrt{ \frac{\omega_k}{ 2 \omega_T}} e^{-\frac{\beta^2}{4} \frac{\omega_k}{\omega_T}}$. For $T=1~\mu$K, $\omega_T= 2 \pi \times 20.8$~kHz and $\omega_T/\omega_k =2.5$. Hence, lattices and atom distributions that are about $T=1~\mu$K deep or wide, respectively, exhibit quantum features such as tunneling and photon-recoil effects. Conversely, for $T=100~\mu$K, $\omega_T= 2 \pi \times 2.08$~MHz and $\omega_T/\omega_k =250$; for this temperature regime and lattice depth we expect convergence of classical and quantum treatments of the COM.

We obtain the excited-state coefficients $b_n$ by solving the above equations for given lattice parameters over a range of detunings $\delta$ and over a range of initial scaled atom momenta $\beta$. For a given $\beta$, the initial condition for solving the TDSE, Eqs.~(\ref{eq:matrixform}),~(\ref{eq:Hmatrix-O1}), and~(\ref{eq:Hmatrix-1-O1}) or (\ref{eq:Hmatrix-1-O1odd}), is $b_n = 0$, $a_n= \delta_{n,n_0}$ with
$n_0 = {\rm{NINT}} (\beta/2)$ and $p_0 = \hbar k (\beta - 2 \, n_0)$, where the function ${\rm{NINT}}(x)$ is the integer closest to $x$. The range of $\beta$ is adapted to temperature; we use ranges
$\beta \in [-10,10]$ for $T=1~\mu$K and $\beta \in [-80,80]$ for $T=100~\mu$K.
For the time-dependence of the drive, $\Omega(t)$, we use Gaussian pulses with durations of tens to hundreds of $\mu$s.
The drive pulse $\Omega(t)$ peaks at time $t_p/2$, and the TDSE is integrated from $t=0$ to $t_p$.
The target-state population is evaluated at $t=t_p$ and averaged over the Maxwell momentum distribution, yielding
\begin{equation}
P_b (\delta)  =  \int \sum_n |b_n(\beta,\delta, t_p)|^2 f (\beta)  d\beta \, .
\label{quasiE}
\end{equation}
In the calculation, the duration $t_p$ is set sufficiently large such that the population $P_b (\delta)$ becomes independent of $t_p$. The population $P_b (\delta)$ is equivalent to experimental spectra in which the Rydberg target state population is measured as a function of lattice modulation frequency, and is therefore a main result of this paper.

\section{Band structure model}
\label{sec:bs}

The above described method of finding the spectrum $P_b (\delta)$ has the advantage that drive pulses $\Omega(t)$ of any time dependence can be modeled. Further, the spectrum usually exhibits some Fourier broadening, which occurs when the drive pulse is substantially shorter than the inverse Rabi frequeny ($t_p \lesssim \pi /\Omega$), or a certain amount of saturation broadening, which occurs when the pulse area exceeds $\pi$ (that is, if $t_p \Omega \gtrsim \pi$). Solving the TDSE is a good way to account for these effects.

For an approximate and fast solution, we use perturbation theory (Fermi's golden rule, FGR) to obtain transition probabilities between the Bloch states of the initial and target Rydberg levels in their respective optical lattices (which have identical periods and Brillouin zones). The transition probabilities are averaged over the Maxwell momentum distribution along the same lines as described in Sec.~\ref{subsec:Max}. These results are a good approximation if the transition is not saturated ($t_p \Omega \ll \pi$) and if COM coherences are not important, as discussed further below. In the following, the FGR method is briefly outlined.

Diagonalizing the Hamiltonian in Eq.~(\ref{eq:Hmatrix-O1}) for both Rydberg levels
$\vert 1 \rangle$ and $\vert 2 \rangle$ separately, we find the initial and target Bloch states
\begin{eqnarray}
\vert \psi_{A,i}(p_0) \rangle & = &  \sum_{n} a_{i,n,p_0} \vert p_0 + 2 n \hbar k\rangle  \nonumber \\
\vert \psi_{B,i'}(p_0) \rangle & = &  \sum_{n} b_{i',n,p_0} \vert p_0 + 2 n \hbar k\rangle  ,
\end{eqnarray}
where $A$ and $B$ stand for the initial and target Rydberg levels, respectively, $i$ and $i'$ are band indices, and the quasimomentum
$p_0 \in \, [-\hbar k, \hbar k]$. We typically obtain the Bloch states and their respective band energies, $E_a(i,p_0)$ and $E_b(i',p_0)$, on an equidistant grid of 200 $p_0$-values. The squares of the transition matrix elements for a time-independent Rabi frequency $\Omega$ in Eqs.~(\ref{eq:Hmatrix-1-O1}) and~(\ref{eq:Hmatrix-1-O1odd}) are
\begin{eqnarray}
\vert V_e (i',i,p_0) \vert^2  & = & \frac{\Omega^2}{4} \vert \sum_{n} (b^*_{i',n+1} + b^*_{i',n-1})a_{i,n} \vert^2 , \nonumber \\
\vert V_o (i',i,p_0) \vert^2  & = & \frac{\Omega^2}{4} \vert \sum_{n} (b^*_{i',n+1} - b^*_{i',n-1})a_{i,n} \vert^2
\label{eq:fgrrate}
\end{eqnarray}
for even- and odd-parity transitions, respectively, and with $i'$ denoting a target- and $i$ an initial-level band index. The $p_0$-dependence of the $a$- and $b$-coefficients is suppressed for brevity. Following FGR for the case of a harmonic drive, the transition rate from the initial Bloch state $i$ into the target Bloch state $i'$ then is
\[
R(i',i, p_0, \delta) = \frac{2 \pi}{\hbar} \vert V (i',i,p_0) \vert^2 \rho(\Delta E) , \]
with energy detuning $\Delta E$ of the lattice-modulation drive from the transition energy between the initial and target Bloch states, $\Delta E (i',i,p_0,\delta) = E_b(i',p_0) - E_a(i,p_0) - \hbar \delta$. For the energy density of states $\rho(\Delta E)$ we use a Gaussian, $\rho(\Delta E) =  \frac{1}{\sqrt{2 \pi} \sigma_E} \exp(- \Delta E^2 / (2 \sigma_E^2))$.
There, the spectral width of the drive is chosen in the range 1~kHz $\lesssim \sigma_E/h \lesssim$ 100~kHz, in accordance with the Fourier widths for our typical drive-pulse durations.

If the drive pulse was a square pulse, the (non-saturated) FGR transition probability would be $P(i',i, p_0, \delta)=R(i',i, p_0, \delta) \, t_p$, with pulse duration $t_p$. The transition rate depends on time if the Rabi frequency $\Omega$ in Eq.~(\ref{eq:fgrrate}) depends on time. In the examples discussed in Sec.~\ref{sec:results}, the Rabi frequency $\Omega(t)$ has a Gaussian time dependence. In that case, $P(i',i, p_0, \delta)$ is given by the integral of $R(i',i, p_0, \delta,t)$ over the duration of the pulse.

It is then assumed that the initial COM states have a normalized thermal probability distribution for temperature $T$,
\[ W(n_0, p_0) \propto \exp(-(p_0 + n_0 \hbar k)^2/(2m k_B T)) , \]
with $p_0 \in \, [ - \hbar k, \hbar k ]$ and integer $n_0$.
For a sudden lattice turn-on, the thermally populated momentum base states $\vert p_0 + 2 n_0 \hbar k \rangle$ are projected into the basis of Bloch states, where the expansion coefficients $a_{i,n_0,p_0}$ are known from the diagonalization of the Hamiltonian in Eq.~(\ref{eq:Hmatrix-O1}). The FGR spectrum is then obtained from
\begin{equation}
Q_b (\delta) = \sum_{i',i,p_0} \left[ P(i',i, p_0, \delta) \sum_{n_0} W(n_0, p_0) \vert a_{i,n_0,p_0} \vert^2 \right] .
\end{equation}

The quantity $Q_b(\delta)$ is the FGR transition probability per atom, averaged over the initial thermal distribution of atoms over free-particle momentum states, for the case that the amplitude-modulated POL is suddenly turned on. This corresponds with the analysis performed in Sec.~\ref{sec:SE}.

\section{Semi-classical model}
\label{sec:sc}

In the limit of temperatures and potential depths equivalent to energies much larger than the lattice recoil energy, $\hbar \omega_k /4$, semi-classical and quantum results should converge in certain aspects. To explore classical-quantum correspondence in the system, we use a model~\cite{Moore2015,MooreMagic} in which the COM dynamics is treated classically by solving Newton's equations with a 6th-order Runge-Kutta routine. As the atoms move, the internal-state dynamics in the Rydberg-state space $\left\{\vert 1 \rangle, \vert 2 \rangle\right\}$ is propagated quantum-mechanically, taking the explicit and implicit time dependence of the Rabi frequency into account. For even-parity transitions, for instance, the Rabi frequency in the atomic frame follows $\Omega(t) \cos(2 k z(t))$, where $\Omega(t)$ is the real-valued, positive Rabi frequency from Eqs.~(\ref{Master1}) and~(\ref{Master-Moment-O1}), which is a Gaussian pulse in our examples, and $z(t)$ is the classical COM position of the atom. Doppler effects arise from the sign flips of the cosine function that occur when the atoms pass through the inflection points of the cosine function. The sign flips are equivalent to phase jumps of $\pi$, which, when occurring at semi-regular time intervals in the moving frame of a hot atom moving through the lattice, cause a Doppler effect similar to the usual Doppler effect (see discussion in Sec.~\ref{subsec:g2b}). The internal evolution also depends on the detuning between the atomic transition and the POL AM modulation frequency, $\delta$. In cases of non-magic lattices ($V_1 \ne V_2$) in the frame of reference of a moving atom with classical COM position coordinate $z(t)$, the detuning further depends on position; in that case, the detuning for the internal quantum evolution is $\delta_c = \delta + (V_2 - V_1) \cos(2 k z(t))$.

As the states $\vert 1 \rangle$ and $\vert 2 \rangle$ have generally different POL amplitudes $V_1$ and $V_2$, a procedure is needed to compute a classical force. Here, we average the classical force over the two lattice potentials for the states $\vert 1 \rangle$ and $\vert 2 \rangle$, using the time-dependent quantum-mechanical probabilities of the atom being in
$\vert 1 \rangle$ or $\vert 2 \rangle$ as weighting factors. This method does not account for COM quantum effects such as tunneling, dispersion, and state-dependent wave-packet splitting. This deficiency ties into the overall failure of the semi-classical description at low energies, and in certain other cases. Nevertheless, over wide swaths of parameter space the semi-classical model is quite successful. It is noted that in magic lattices  ($V_1 = V_2$) the two coupled internal states have identical potentials, in which case the classical force simplifies to $F(z(t)) = 2 k V_1 \sin(2 k z(t))$.

In accordance with the quantum models, in the semi-classical description initial atom velocities $v(t=0)$ are drawn from the Maxwell distribution given in Sec.~\ref{subsec:Max}, and initial positions $z(t=0)$ are random. The semi-classical spectrum, $K_b(\delta)$, is given by the transition probability at the end of the interaction time $t_p$, averaged over a classical thermal ensemble of initial atoms. In this work, we average over 10,000 to 100,000 trajectories.

\section{Results}
\label{sec:results}

In general, it is desired to strike a balance between observing a transition with high signal-to-noise ratio and avoiding saturation broadening. Therefore, in our calculations we use Rabi frequencies that, for a given drive pulse shape and duration, yield a $\pi$ pulse area for an atom at a COM location $z$ where the Rabi frequency is maximal, $\int \Omega(t) dt = \pi$. For our Gaussian pulses, $\Omega(t) =\Omega_0 e^{-(t-t_p/2)^2/\tau_0^2}$.  In the cases discussed below we quote the full width at half maximum of the pulse, which is $\tau_{\mathrm{FWHM}}=2\tau_0\sqrt{\ln (2)}$, and the corresponding utilized value of $\Omega_0$ that leads to the $\pi$ pulse.

\begin{figure*}[t]
\centerline{\includegraphics[width=15cm]{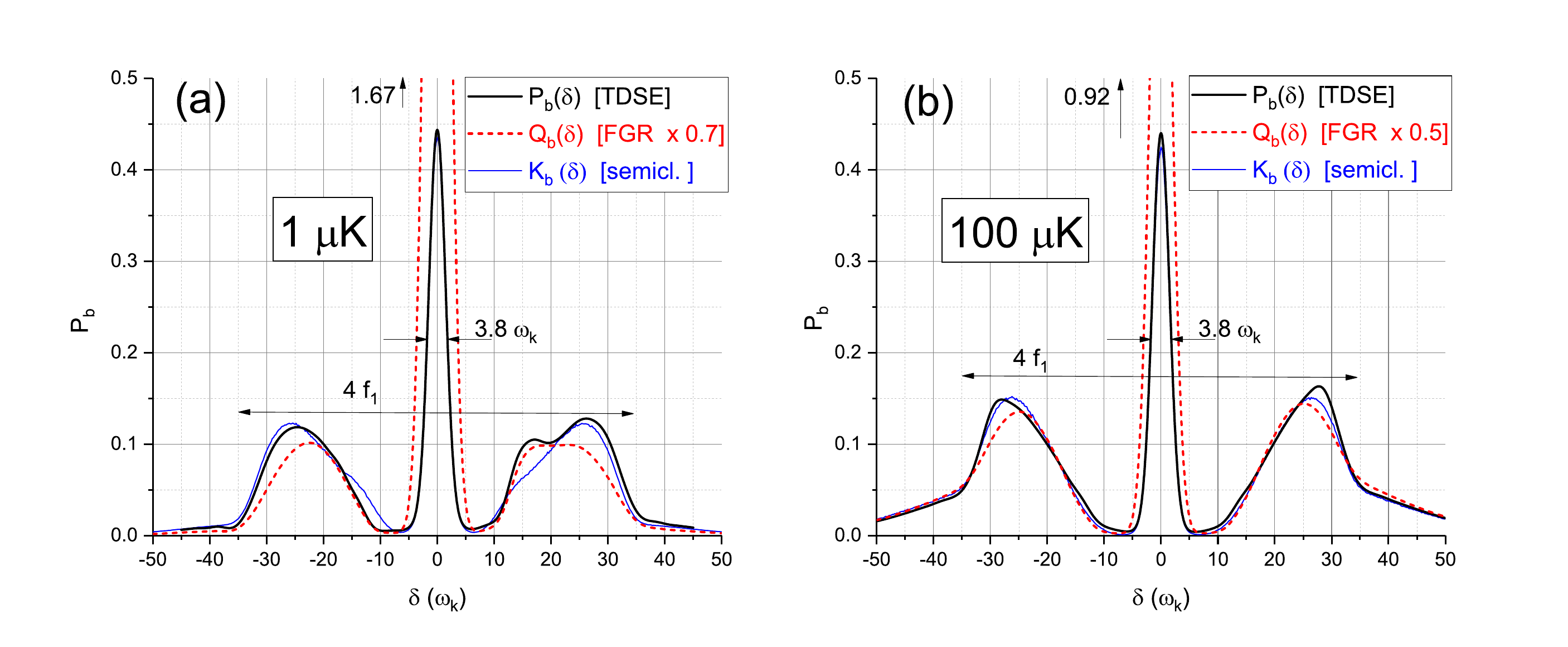}}
\caption{(Color online) POL modulation spectra $P_b$ (TDSE), $Q_b$ (FGR) and $K_b$ (semi-classical) for $V_1 = V_2 = 2 \pi \times 1.25$~MHz, $\tau_{\mathrm{FWHM}} = 20~\mu$s, and $T=1~\mu$K (a) and $100~\mu$K (b).}
\label{fig:g2}
\end{figure*}

\subsection{Structure of POL amplitude-modulation spectra}
\label{subsec:g2a}

The range of pulse durations of interest is between $1~\mu$s and $1~$ms, as this range is accessible given the lifetimes of typical Rydberg states~\cite{GallagherBook}. Within this range, we consider a pulse short if
its Fourier bandwidth suffices to resolve the vibrational sidebands $\Delta \nu \ne 0$ in the spectrum from the central $\Delta \nu = 0$ peak, but the bandwidth is too large to resolve the recoil energy and anharmonic effects of the band structure of the COM (see Fig.~\ref{fig:g1}). Conversely, a pulse is considered long if it resolves the anharmonicity-induced quantum structures in the vibrational sidebands.  For two selected cases of short and long pulses, we will discuss spectra for several representative lattice depths and COM atom temperatures. Our initial discussion is focused on the TDSE and semi-classical  models.

In Fig.~\ref{fig:g2} we show POL modulation spectra $P_b(\delta)$, $Q_b(\delta)$ and $K_b(\delta)$, obtained by solving the TDSE, by computing FGR transition probabilities between Bloch states, and by performing semi-classical simulations, respectively. The transitions are even-parity and are driven by Gaussian field pulses with $\tau_{\mathrm{FWHM}} = 20~\mu$s, corresponding to $\tau_0 = 12.0~\mu$s. Our calculations are for $\omega_k = 8.300$~kHz, corresponding to $^{85}$Rb atoms in a POL formed from two counter-propagating beams of 1064~nm wavelength, and lattice amplitudes $V_1 = V_2 = 2 \pi \times 1.25$~MHz (a case of a magic lattice). These conditions are similar to experimental work in~\cite{Moore2015,MooreMagic}. The spectra exhibit the lowest-order allowed vibrational sidebands, $\Delta \nu = \pm 2$, as well as a Doppler-free central band $\Delta \nu = 0$. Since the COM oscillation frequency of the atoms at the bottoms of the wells is $2 k \sqrt{\hbar V_1/m} = 2 \pi \times 144$~kHz, the frequency separation between the $\Delta \nu = \pm 2$ vibrational sidebands is about 69~$\omega_k/(2 \pi)$. Inspecting Fig.~\ref{fig:g2}, this corresponds to the separation between the outer fringes of the sidebands. The average separation is somewhat less, because atoms that are thermally excited into vibrational states of the COM motion above the vibrational ground state have a smaller oscillation frequency, leading to less separation. Anharmonicity-induced substructures are not resolved due to the FWHM width of the spectral density of the pulse, which is $3.77 \omega_k = 2 \pi \times 31.3~$kHz (in the case of low saturation, for the given pulse length). Also, higher-order sidebands,  $\Delta \nu = \pm 4, \pm6 , ...$ are too broad and weak to be observed.

For measurement and metrology purposes, the Doppler-free peak for $\Delta \nu = 0$ is of particular interest, which in
Fig.~\ref{fig:g2} has a FWHM of $\lesssim 4\omega_k$. This is in good agreement with the low-saturation Fourier width of the Gaussian pulse, 3.77$\omega_k$. Note that the presented case is for a $\pi$-pulse, and therefore the central peak is slightly saturation-broadened.

\subsection{Doppler-free spectroscopy in modulated POL}
\label{subsec:g2b}

The origin of the Doppler-free nature of the central peak becomes clear from the following semi-classical analysis. The classical COM of an atom oscillating within the center portion of a lattice well is approximately harmonic, $z(t) = z_1 \cos(2 \pi f_1 t)$, with COM oscillation frequency $f_1 = k \sqrt{\hbar V_1/m} / \pi$ and COM oscillation amplitude $z_1 \lesssim \lambda/8$. Using the Jacobi-Anger expansion, the time dependence of the Rabi frequency in the frame of the atom is given by
\begin{eqnarray}
\Omega_a(t)  & = & \Omega(t) \cos(2 k z(t)) \nonumber  \\
~ & = & \Omega(t) [ J_0 (2 k z_1) \cos(2 k z_1) \nonumber  \\
~ & + & 2 \sum_{p=1} (-1)^p J_{2p} (2 k z_1) \cos(4 \pi p f_1 t) ] .
\label{eq:FM2}
\end{eqnarray}
Here, the explicit time dependence is contained in the Gaussian envelope function $\Omega(t) =  \Omega_0 e^{-(t-t_p/2)^2/\tau_0^2}$ with constant $\Omega_0$. The implicit time dependence in Eq.~(\ref{eq:FM2}) arises from the atomic motion, $z(t)$, and is contained in the Fourier series in the square brackets. The atomic motion generates Fourier components of the drive at even multiples of $f_1$, which correspond with the vibrational signals for $\Delta \nu = 0$ and $ \pm 2$ in  Fig.~\ref{fig:g2}. For the case of Fig.~\ref{fig:g2}, $\Omega(t)$ is a slowly-varying real-valued envelope function with a duration of a few COM oscillation periods of the atoms in the wells. The $\Delta \nu = 0$ and $\pm 2$ components are resolved in Fig.~\ref{fig:g2} because the Fourier width of the envelope $\Omega(t)$ is less than the frequency separation $2 f_1$ between the components. This basic interpretation applies if a substantial fraction of atoms is trapped within the approximately harmonic regions of the POL wells, and if the spectrum of the pulse envelope $\Omega(t)$ is sufficiently narrow to resolve the vibrational sidebands. The treatment based on the band structure and FGR, visualized in Fig.~\ref{fig:g1}, yields equivalent conclusions. Importantly, in Eq.~(\ref{eq:FM2}) it is evident
that the atom velocity plays no direct role in the spectrum. In particular, in magic lattices as in Fig.~\ref{fig:g2} the cental peak $\Delta \nu = 0$ is Doppler-effect-free and is unaffected by the anharmonicity of the lattice, making it ideal for high-precision spectroscopy of Rydberg-atom transitions.

It is noted that Eq.~(\ref{eq:FM2}) resembles the spectrum of frequency-modulated fields and the spectrum seen by optically driven ions oscillating in ion traps. These similarities can be born out more clearly in a photon picture of POL modulation spectroscopy, in which the atoms scatter a photon from one lattice-field mode into a counter-propagating field mode, where
the mode frequencies differ by the POL modulation frequency (which equals the atomic-transition frequency and is orders of magnitude larger than $f_1$ and $\omega_k$). The re-scattering is a stimulated process effected by the $A^2$-term of the atom-field interaction. A detailed analysis of this picture is not of interest in the present paper.

Spectral broadening akin to the usual Doppler effect arises from atoms traversing over many lattice wells. Again, a semi-classical picture is well-suited to explain this effect. The phase of the drive term in Eq.~(\ref{Master1}), $\Omega(t) \cos(2 k z)$, undergoes a jump of value $\pi$ at every inflection point of the lattice potential. For a hot atom moving at constant velocity $v$ through the lattice, in the reference frame of the moving atom the phase
of the drive field follows a step-function that is centered around the linear function
$\phi (t)\approx 2 \pi z(t) / (\lambda/2)$ $= 2 k v t$, equivalent to a Doppler shift of $2 k v$. This resembles the Doppler shift of stimulated Raman scattering between counter-propagating beams. It is noted, however, that POL modulation spectroscopy is fundamentally different from Raman spectroscopy, because it employs a first-order $A^2$-process and not a second-order ${\bf{A}} \cdot {\bf{p}}$-process.

\subsection{Temperature insensitivity of spectroscopy in modulated POL}
\label{subsec:g2c}

If the lattice depth $2 \hbar V_1 \gtrsim k_B T$, the initial temperature has only a minor effect on the spectrum. This is seen clearly in Fig.~\ref{fig:g2}, where the spectra for $T=1~\mu$K and $100~\mu$K are quite similar. The heights and widths of the central peaks are near-identical in both cases. The temperature insensitivity results from the fact that atoms initially located away from a lattice minimum gain considerable potential energy at time $t=0$, when the lattice is suddenly turned on. If $2 \hbar V_1 \gtrsim k_B T$, the initial potential energy dominates the initial kinetic energy, and the distributions of oscillation amplitudes $z_1$ are not very different (in the semi-classical model). Hence, for $2 \hbar V_1 \gtrsim k_B T$ the signal strengths in the vibrational sidebands, $\Delta \nu = \pm 2$, are not very temperature-dependent.
In the TDSE and band-structure models, the temperature insensitivity follows from the fact that the projection of a thermal ensemble of plane waves of the COM motion into the Bloch-state basis yields similar distributions as long as the potential energy is larger than the thermal energy of the ensemble prior to projection, leading to the same condition,  $2 \hbar V_1 \gtrsim k_B T$.

\subsection{Quantum-classical correspondence}
\label{subsec:g2d}

Comparing the exact TDSE and the semi-classical spectra in Fig.~\ref{fig:g2}, it is seen that the semi-classical model does quite well, even at a quantitative level. A leading deviation between TDSE and semi-classical results is that the vibrational sideband structure (for magic transitions) is symmetric in the semi-classical calculation, whereas it is asymmetric in the TDSE result. The symmetry in the semi-classical case directly follows from Eq.~(\ref{eq:FM2}), where positive- and negative-frequency components of the drive have identical amplitude. After dropping the harmonic-COM approximation made in Eq.~(\ref{eq:FM2}), this symmetry still holds. It is also noted that in the semi-classical model the red- and blue-detuned transitions between the Rydberg states $\vert 1 \rangle$ and $\vert 2 \rangle$ have no recoil effect on the classical COM motion, in accordance with the
perfect symmetry between blue- and red-detuned vibrational sidebands in the semi-classical results.

In contrast, in both quantum treatments (TDSE and FGR) the spectra are non-symmetric. Using, for the sake of clarity, the notion of a harmonic COM motion, the $\hat{z}^2$-term in the expansion of the cosine in the drive term, $\Omega(t) \cos(2 k \hat{z})$, causes (most of) the $\Delta \nu = \pm 2$ vibrational sidebands in the spectrum.
While the relevant matrix elements between COM states, $\langle \nu -2 \vert \hat{a}^2 \vert \nu \rangle$ and $\langle \nu \vert \hat{a}^{\dagger 2} \vert \nu - 2 \rangle$, are symmetric, the population difference between the initial vibrational COM states $\vert \nu - 2 \rangle$ and $\vert \nu \rangle$ reduces the red-shifted sideband relative to the blue-shifted one by a Boltzmann factor of about $\exp(-2 h f_1 / (k_B T_{\mathrm{eff}}))$, where $T_{\mathrm{eff}}$ is an effective COM temperature after the sudden transfer of the atoms into the lattice.  Generally, the vibrational spectral sidebands are the more asymmetric the lower the temperature and the shallower the lattices are.

\subsection{Exact TDSE solution versus perturbative model}
\label{subsec:g2e}

In Fig.~\ref{fig:g2}, the perturbative FGR-model based on the band structure reproduces the TDSE results fairly well.
As our drive pulses are near saturation of the transition, the FGR model generally overestimates the transition probabilities. This is qualitatively adjusted by scaling the FGR probabilities by a factor $<1$.
The FGR result exhibits asymmetries similar to those in the exact TDSE solution, as expected (see Sec.~\ref{subsec:g2d}).
There are, however, deviations between the shapes of the FGR and the TDSE results. This is attributed to the fact that the FGR model does not account for any coherent transients, which can cause strong effects in the case of short drive pulses.
This includes COM transients, which are generated by the sudden transfer of the atoms from free space into the lattice, which initiates a COM wave-packet motion. Due to the position dependence of the drive term, $\Omega(t) \cos(2kz)$, the COM wave-packet maps onto a transient signal in the internal-state dynamics (which is our observable). The transients are the most pronounced if the pulse duration is on the order of the harmonic period of the COM motion, $1/f_1$.  The cases discussed in the present and even more so in the next subsection are in this regime.

While our FGR model cannot describe the transients, the transients are reproduced in large parts in the semi-classical model. This is not unexpected, because the semi-classical model incorporates a classical approximation of COM dynamics, the origin of the transients. Generally, the transients-related deviations between the FGR and the TDSE and semi-classical results are most visible in shallow lattices and at high temperature. In Fig.~\ref{fig:g2} the POL is deep enough that the deviations between the FGR lineshapes and the TDSE/semi-classical lineshapes are still fairly minor.

\begin{figure*}[t]
\centerline{\includegraphics[width=15cm]{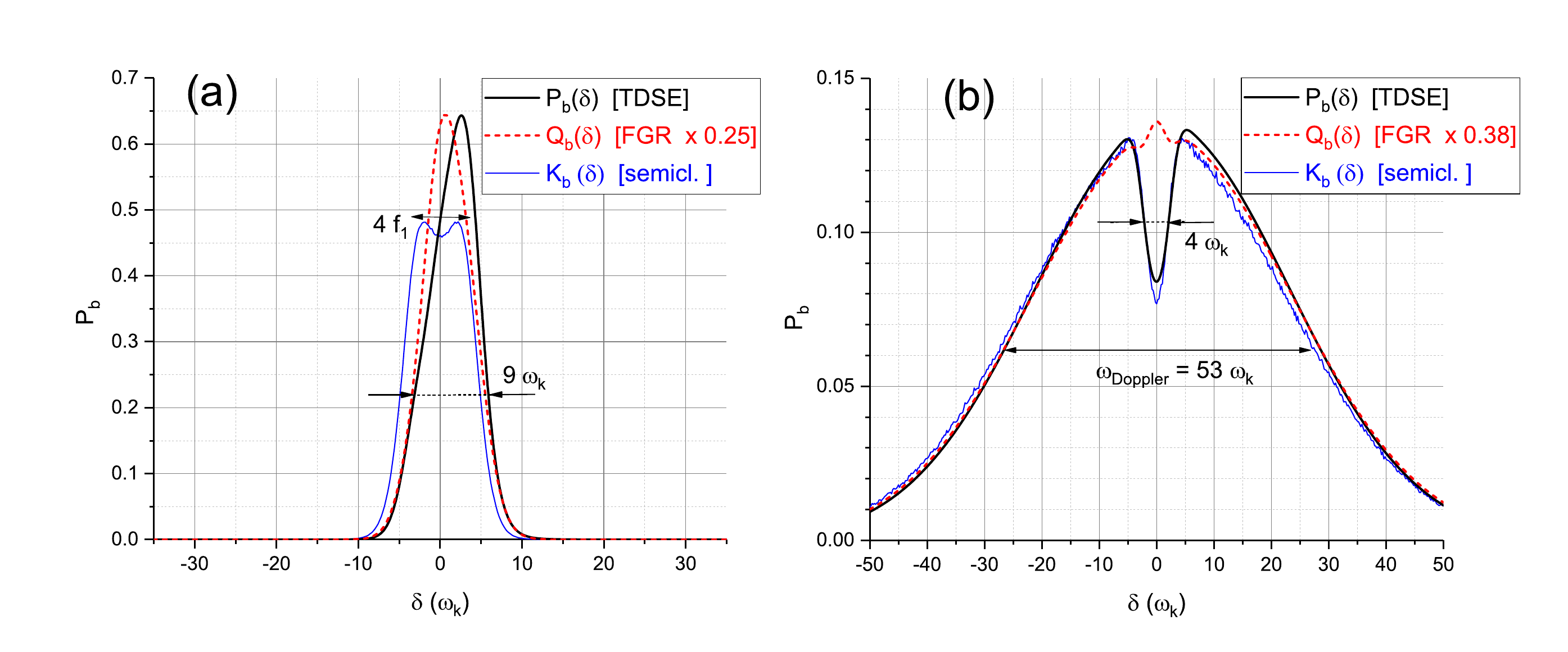}}
\caption{(Color online) POL modulation spectra $P_b$ (TDSE), $Q_b$ (FGR) and $K_b$ (semi-classical) for $V_1 = V_2 = 2 \pi \times 12.5$~kHz, $\tau_{\mathrm{FWHM}} = 20 \mu$s, and $T=1~\mu$K (a) and $100~\mu$K (b).
 } \label{fig:g3}
\end{figure*}

\subsection{POL amplitude-modulation spectra in shallow lattices}
\label{subsec:g3}

Spectroscopic data often yield better results in less deep lattices due to a reduction in residual AC shifts, reduced photo-ionization losses, etc. In Fig.~\ref{fig:g3} we show results for the same parameters as in Fig.~\ref{fig:g2}, with the exception that the lattice is only 1/100-th as deep. In this case, the POL has only two tightly-bound bands and an $f_1$-value of only $14.4~$kHz, corresponding to a round-trip time of $70~\mu$s.

At $T=1~\mu$K, quantum and semi-classical results still agree fairly well. The asymmetry of the TDSE result is much larger in the shallow lattice (Fig.~\ref{fig:g3}) than it is in the deep one (Fig.~\ref{fig:g2}). The asymmetry is somewhat reproduced in the FGR result, whereas the semi-classical spectra are perfectly symmetric and show a hint of resolving the $\Delta \nu = \pm 2$ sidebands.  These trends in overall agreement and asymmetry behavior are expected from the discussions in the previous sections. Deviations between TDSE and FGR results are again attributed to the effects of COM transients on the internal-state dynamics.

At $T=100~\mu$K, the TDSE and semi-classical results (Fig.~\ref{fig:g3}~(b)) agree very well, as the atoms are comparatively hot, leading to essentially classical COM dynamics. It is noteworthy that both results exhibit a central dip with a width near the Fourier limit. The dip is attributed to a rotary-echo behavior, which is most significant when $\tau_0 \sim 1/(4 f_1)$, as is the case in Fig.~\ref{fig:g3}~(b). To explain the rotary-echo effect, we consider a drive-pulse duration on the order of $1/(4 f_1)$. In that case, the classical motion of a large fraction of atoms covers about one half of the lattice period, over the duration of the drive pulse. Due to the cosine-dependence of the Rabi frequency, the atoms spend similar amounts of time in spatial regions with positive and with negative-valued Rabi frequencies. The resultant rotary-echo effect~\cite{Solomon1959,Raitzsch2008,Younge2009,Thaicharoen2017} can drastically reduce the excitation probability, as is most clearly seen Fig.~\ref{fig:g3}~(b).

The FGR model entirely fails to produce the echo-induced dip seen in the exact TDSE solution and the semi-classical result. This is due to the fact that our FGR model does not account for COM wave-packet dynamics and COM coherences at all. Any rotary-echo effects induced by COM motion and position-dependent sign flips of the Rabi frequency can therefore not be described with the FGR model. The absence of a dip in the FGR-spectrum in Fig.~\ref{fig:g3}~(b) signifies that in certain cases the rotary-echo effect is critical in understanding POL modulation spectra.

We notice that width of the peak in the lower-temperature case (Fig.~\ref{fig:g3}~(a)) is substantially larger than the echo-induced dip in the high-temperature spectrum (Fig.~\ref{fig:g3}~(b)). Thus, suitable combinations of lattice depth and drive-pulse duration yield high-temperature spectra that are more suitable for high-precision spectroscopy than their low-temperature counterparts.

\subsection{POL amplitude-modulation spectra at long interaction times}
\label{subsec:g4}

\begin{figure}[h]
\centerline{\includegraphics[width=8cm]{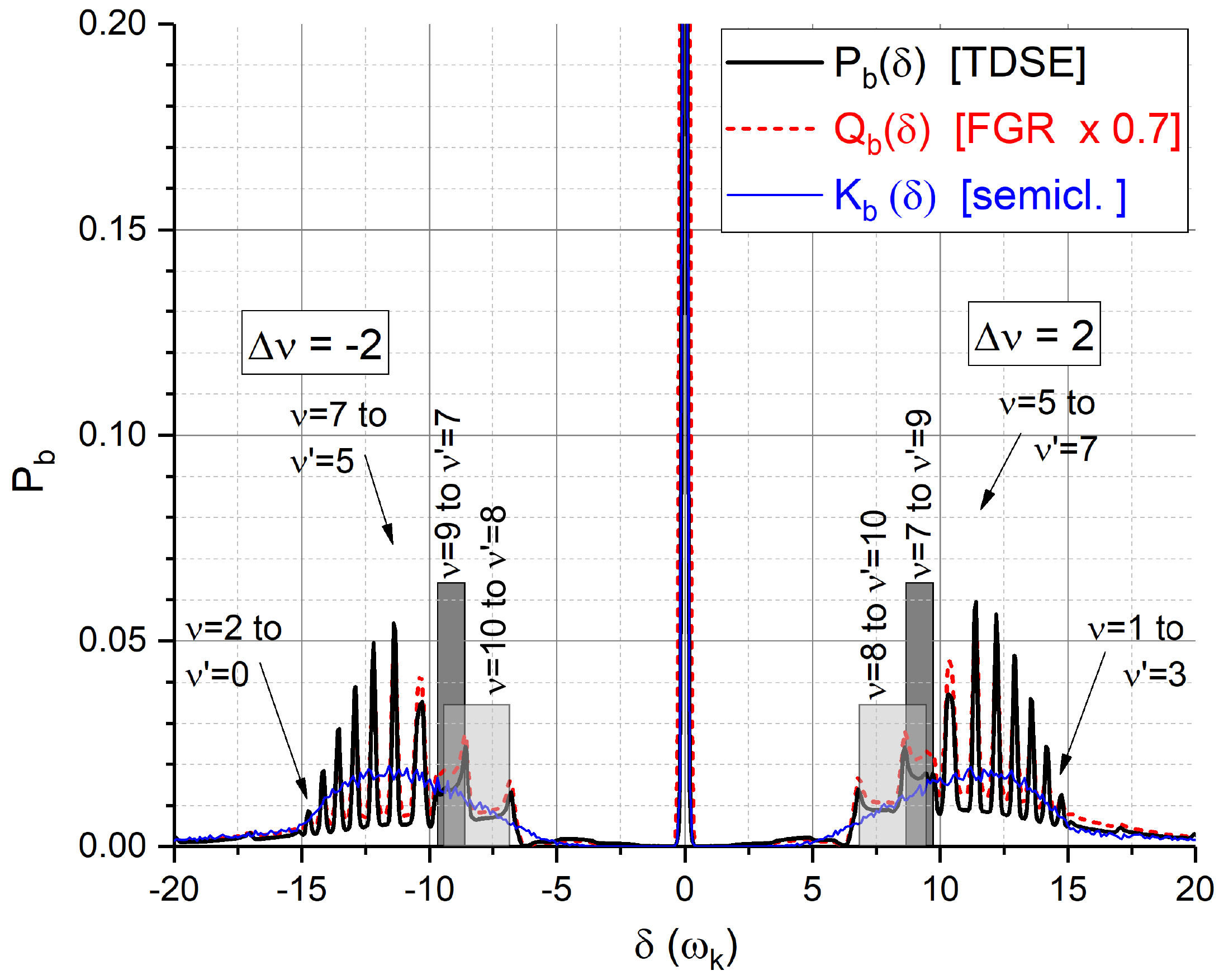}}
\caption{(Color online) POL modulation spectra $P_b$ (TDSE), $Q_b$ (FGR) and $K_b$ (semi-classical) for $V_1 = V_2 = 2 \pi \times 250$~kHz, $\tau_{\mathrm{FWHM}} = 400 \mu$s, and $T=10~\mu$K.
 } \label{fig:g4}
\end{figure}

A prominent prospect of using POL modulation spectroscopy is to perform high-precision measurements with cold, long-lived circular-state Rydberg atoms. To that end, we now consider several cases of moderately-deep magic and non-magic POLs, and drive pulses with a FWHM field-pulse length of 400~$\mu$s. While this is still about an order of magnitude short of typical circular-state lifetimes in a 300\,K thermal-radiation environment, it suffices for a discussion of quantum structures of COM vibrational side-bands, and of vibrationally-resolved Doppler-free spectroscopy in lattices that are non-magic. Here we choose a 10~$\mu$K temperature, a value attainable in rapid-cycle optical-molasses cooling~\cite{MetcalfBook}.

In Fig.~\ref{fig:g4} we consider an even-parity magic transition in a moderately-deep POL with $V_1=V_2=250$~kHz. As before, the Rabi frequency is chosen such that atoms pinned at a spatial Rabi-frequency maximum experience a $\pi$-pulse. It is seen that all models produce a Fourier-limited central peak, with virtually no signal background at detunings $\delta \lesssim 5 \omega_k$.

Since in Fig.~\ref{fig:g4} the drive pulse is much longer than in Figs.~\ref{fig:g2} and~\ref{fig:g3}, the ratio between pulse length and vibrational period is much larger ($\tau_{\mathrm{FWHM}} f_1 = 26$), and the Rabi frequency is smaller. Therefore, the case in Fig.~\ref{fig:g4} is deeper within the validity range of FGR perturbation theory. As a result, in Fig.~\ref{fig:g4} the agreement between the exact solution of the TDSE and the FGR approximation is quite good. Both quantum models show a substructure of the
$\Delta \nu = \pm 2$ vibrational sidebands that arises from level shifts of the excited COM levels. The reduced Fourier width afforded by the 400-$\mu$s long pulse length allows for the observation of the $\nu-$resolved substructure, with each sub-peak of the sidebands characterized by a single $\nu-$value.
The vibrational splitting of the $\Delta \nu = \pm 2$ sidebands arises from the anharmonicity of the wells and is, unsurprisingly, well-represented by both the TDSE and the FGR-band-structure models.

Due to the softening of the POL potential near its maxima, the $\Delta \nu = \pm 2$ transitions for larger $\nu$-values occur at smaller absolute values of the detuning $\delta$. Defining $\nu_{\mathrm{min}}$ as the minimum of the coupled states
$\nu$ and $\nu'$, it is seen in Fig.~\ref{fig:g4}
that for $\nu_{\mathrm{min}} \lesssim 4$ the change in transition frequency is proportional to $\nu_{\mathrm{min}}$. This trend is easily confirmed by considering the effect of the lowest-order non-linear correction of the trapping potential, which is $\propto - \hat{z}^4$, on the transition frequencies within the $\Delta \nu = \pm 2$ sidebands. We also see that the lines for larger $\nu_{\mathrm{min}}$ are broadened according to the widths of the corresponding lattice bands, as visualized by the square boxes in Fig.~\ref{fig:g4} for the $\nu\!=\!7 \rightarrow \nu'\!=\!9$ and $\nu\!=\!8 \rightarrow \nu'\!=\!10$ transitions and their conjugates. Due to the ensemble average that is being taken, the extrema of the band-energy differences, $E_b(p_0)-E_a(p_0)$, which occur at $p=0$ and $p = \pm \hbar k$, produce enhanced signals near the edges of the corresponding spectral features.

The behavior of the strengths of the vibrationally-resolved lines in the quantum results for the
$\Delta \nu\!=\!\pm 2$ sidebands is given by
thermal populations and the squares of the COM matrix elements of $\langle \nu', p_0 \vert \cos(2k\hat{z}) \vert \nu, p_0 \rangle$. For conditions as in Fig.~\ref{fig:g4}, the thermal populations in the lowest few tightly bound states drop off slowly. Over the spatial range of their vibrational COM wave-functions one may approximate $\cos(2k\hat{z}) \approx 1 - \hat{z}^2/2$. Considering the usual expansion of $\hat{z}$ in raising and lowering operators, it is seen that the line strength scales as $(\nu_{\mathrm{min}}+1) (\nu_{\mathrm{min}}+2) \approx \nu_{\mathrm{av}}^2$, where $\nu_{\mathrm{min}}$ is the smaller of $\nu$ and $\nu'$ and $\nu_{\mathrm{av}}$ is the average vibrational quantum number, $(\nu + \nu')/2$. Classically, according to Eq.~(\ref{eq:FM2}) the line strength of the $\Delta \nu\!=\!\pm 2$ transitions is $\propto (J_2(2kz_1))^2$, which for small oscillation amplitudes $z_1$ scales as $z_1^4$, which in turn scales as the square of the quantum-mechanical vibrational quantum number $\nu$. We see that quantum and classical analysis give the same line-strength scaling $\propto \nu_{\mathrm{av}}^2$ for the vibrationally-resolved sub-lines within the $\Delta \nu\!=\!\pm 2$ sidebands. The height-above-base of the lowest few of the sub-lines in Fig.~\ref{fig:g4} clearly follows this trend; the scaling also explains why the $\nu_{\mathrm{min}} = 0$ sub-lines are so weak.

The semi-classical spectrum $K_b(\delta)$ for the $\Delta \nu\!=\!\pm 2$ sidebands in Fig.~\ref{fig:g4} shows excellent qualitative agreement with the $\delta$-averaged quantum spectra, averaged over the $\nu$-quantization in the sidebands. Since in the classical treatment the vibrational COM energy is not quantized, this type of agreement between quantum and classical analyses accords well with our expectations.

\subsection{POL amplitude-modulation spectra in non-magic lattices}

\begin{figure}[h]
\centerline{\includegraphics[width=8cm]{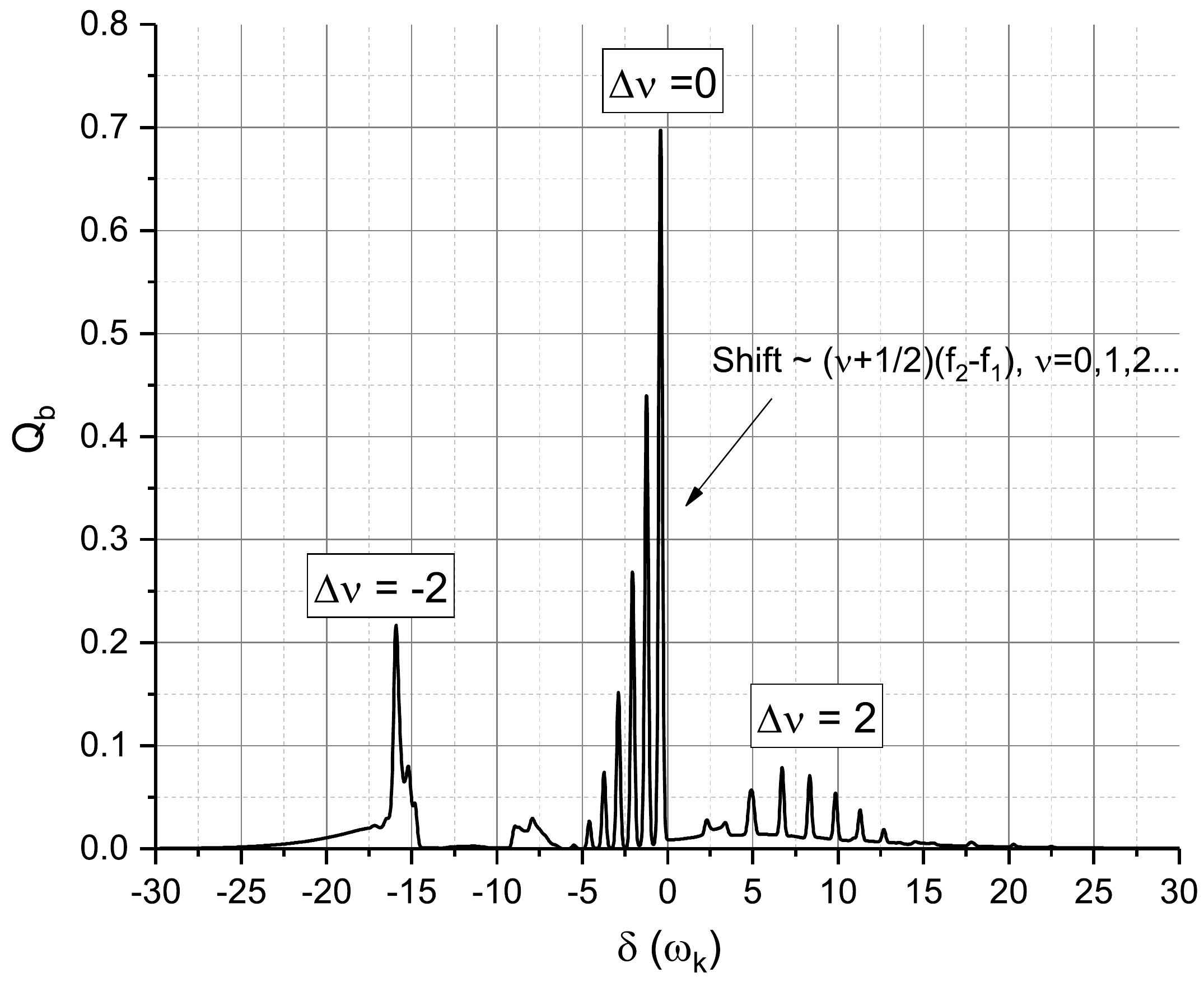}}
\caption{(Color online) POL modulation spectra $P_b$ (TDSE), $Q_b$ (FGR) and $K_b$ (semi-classical) for a non-magic lattice with $V_1 = 2 \pi \times 250$~kHz, $V_2 = 2 \pi \times 200$~kHz, $\tau_{\mathrm{FWHM}} = 400 \mu$s, and $T=10~\mu$K.
 } \label{fig:g5}
\end{figure}

As seen in the previous section, long drive-pulse durations combined with moderately deep POL potentials allow us to resolve vibrational quantization in the $\Delta \nu\!=\!\pm 2$ sidebands. In non-magic lattices, this feature also extends to the $\Delta \nu\!=\!0$ central band. This makes non-magic lattices with a homogeneous lattice-laser intensity distribution suitable for high-precision spectroscopy.
As an example, in Fig.~\ref{fig:g5} we consider a case with parameters identical to those of Fig.~\ref{fig:g4}, except that $V_2 = 0.8 V_1$. Is is seen that the central peak, $\Delta \nu\!=\!0$, splits up into lines
at frequencies $\delta / (2 \pi) \approx (\nu_{\mathrm{min}}+1/2)(f_2-f_1)$, with $\nu_{\mathrm{min}}=0,1,2...$ and using the harmonic approximation. To find the unshifted atomic resonance in a high-precision measurement, one may plot $\delta / (2 \pi)$ against $x = \nu_{\mathrm{min}} + 1/2$. Extrapolation to $x=0$ yields a $y$-intercept that marks the un-shifted atomic resonance. Also, the measured line spacing $f_2-f_1$ allows for a calibration of lattice depths $V_1$ and $V_2$.

Figs.~\ref{fig:g4} and~\ref{fig:g5} show that magic and non-magic POL modulation spectra can, in principle, be modeled in great detail. Thereby, the only narrow spectral feature that is suitable for Doppler-free high-precision spectroscopy and that is, at the same time, insensitive to lattice-depth inhomogeneities is the central ($\Delta \nu =0$) peak in magic lattices. All other Fourier-limited features in Figs.~\ref{fig:g4} and~\ref{fig:g5} are Doppler-free, but exhibit shifts that scale with linear combinations of the COM vibration frequencies, which in turn scale with the square root of lattice power. Assuming that the lattice intensity in the atom-field interaction region can be stabilized to within $\sim 1 \%$, one may expect to be able to resolve the sub-lines within the various vibrational sidebands. In that case, the only free parameters to fit an entire experimental spectrum with a set of calculated spectra will be the lattice depths $V_1$ and $V_2$, and a detuning offset of $\delta$. As the ratio $V_1/V_2$ is known (it only depends on the Rydberg levels, the lattice-laser wavelength and the beam angles~\cite{Dutta2000}), there are only two independent fit parameters. Hence, a two-parameter fit should simultaneously yield a detailed match of all sub-Doppler lines. The two-parameter fit yields  a lattice-depth calibration and a result for the $\delta$-offset. The latter amounts to an accurate measurement of the lattice-free atomic transition frequency that is corrected for lattice-induced transition shifts.

\subsection{POL amplitude-modulation spectra for odd-parity transitions}
\label{subsec:g6}

\begin{figure}[t]
\centerline{\includegraphics[width=8cm]{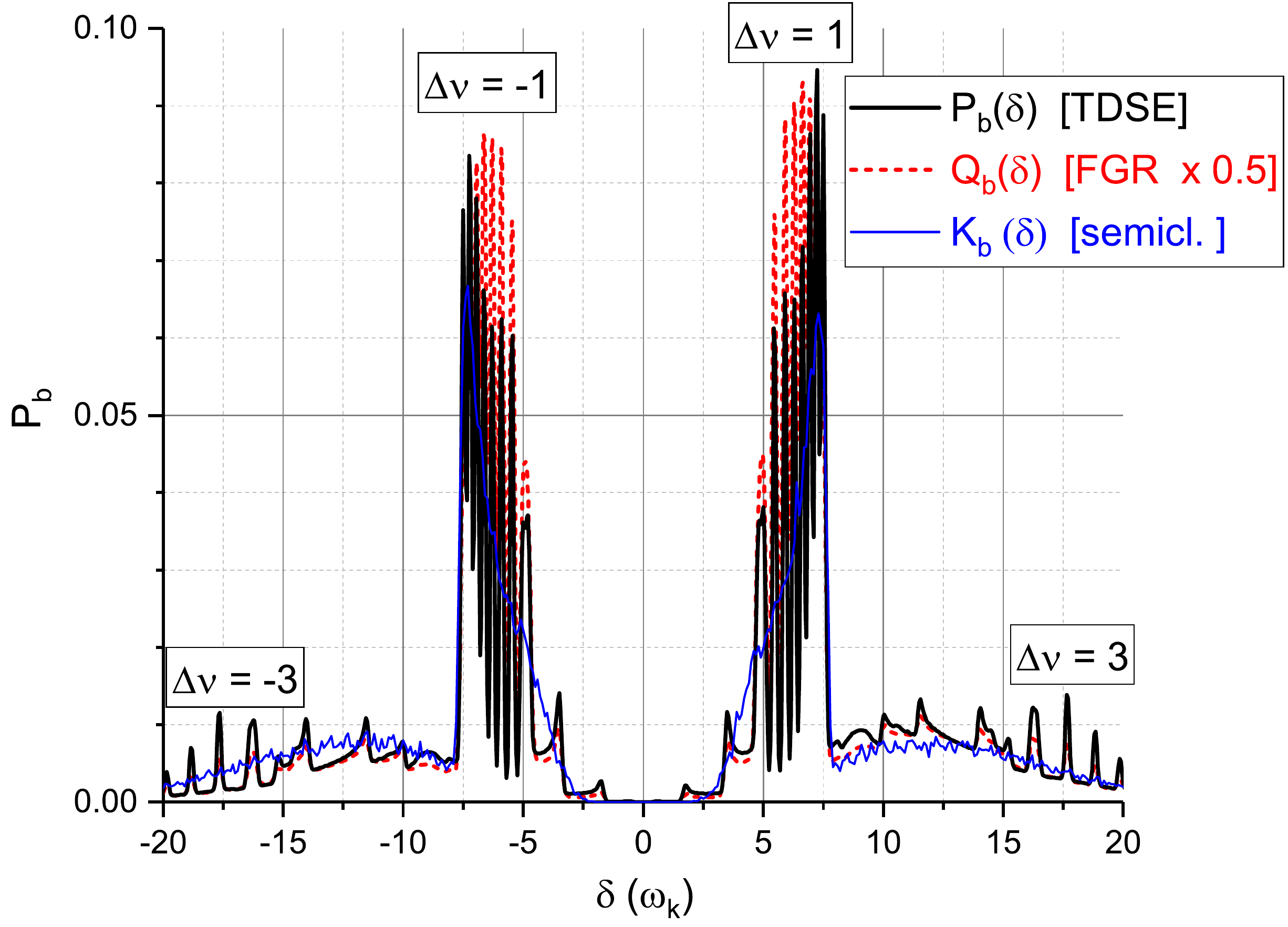}}
\caption{(Color online) POL modulation spectra $P_b$ (TDSE), $Q_b$ (FGR) and $K_b$ (semi-classical) for odd-parity transitions and $V_1 = V_ 2= 2 \pi \times 250$~kHz, $\tau_{\mathrm{FWHM}} = 400 \mu$s, and $T=10~\mu$K.} \label{fig:g7}
\end{figure}

In Fig.~\ref{fig:g7} we finally consider a case of a magic lattice with odd-parity drive. In this case, the vibrational selection rules are $\Delta \nu = \pm1, \pm3, ...$, with corresponding vibrational shifts. Most importantly, the central Doppler-free feature (which in magic lattices is insensitive to lattice-depth variations) is forbidden, and the lowest-order allowed vibrational bands are $\Delta \nu = \pm1$. These are separated in frequency by $2 f_1$. Given a drive pulse with a narrow enough spectrum, the bands are split into vibrational sub-lines due to the anharmonicity of the lattice. For a sufficiently well-controlled lattice intensity, POL modulation spectroscopy of odd-parity transitions could also be employed in high-precision spectroscopy work.

\section{Conclusion}

We have presented three models of modulation spectroscopy of Rydberg atoms
in ponderomotive optical lattices (POL), an application that harnesses the often-ignored $A^2$-term in the atom-field
interaction. Foremost, it has been stressed
that the modulation-induced drive term generates a spatially periodic Rabi frequency with a step-function phase dependence on position. This peculiar phase behavior of the drive enables a novel type of sub-Doppler spectroscopy suitable for applications in high-precision measurement in Rydberg-atom systems. The vibrationally-resolved lines rely on the quantum entanglement between electronic and COM motion, as well as the peculiar couplings afforded by POL modulation.

In our case studies we have used modulation functions with a Gaussian time dependence, because the spectra are free of Fourier sidebands (that would arise in square-pulse drives, for instance). It is, in principle, fairly straightforward to realize modulation drive pulses with a time-dependent Rabi frequency. This can be done by implementing time-dependent POL amplitude modulation with electro-optic fiber modulators.

We have developed a picture of quantum-classical correspondence between the quantum and semi-classical models. We have found a rotary-echo phenomenon that arises from the interplay between the center-of-mass motion of the atoms and the spatial dependence of the modulation-induced Rabi frequency. Our perturbative quantum model fails to reproduce the rotary-echo effect, as expected. The rotary-echo effect can improve spectral resolution in some cases.

A variety of Fourier-limited, Doppler-free vibrational transitions is seen in fairly deep lattices, even at temperatures in the range of 100~$\mu$K. Magic Rydberg-atom optical lattices lead to the most robust spectroscopic structure. However, non-magic lattices with differences on the order of $20\%$ between the lower- and upper-state potential depths should yield equivalent spectroscopic accuracy and precision, when using suitable fitting methods. Ponderomotive transitions can also be driven by spatial ``shaking'' of the lattice; this is possible by phase modulation of a lattice beam (as opposed to amplitude modulation of the entire lattice, the case studied in the present paper).
In any case, POL modulation spectroscopy is expected to yield line widths in the kHz-range, opening venues for Rydberg-atom-based high-precision spectroscopy~\cite{Ramos2017} and quantum simulators~\cite{Nguyen2018}.

\section{Acknowledgments}
This work was supported by NASA (Grant No.NNH13ZTT002N NRA) and NSF (Grant No. Grants No. PHY-1806809).

$^{\dag}$Present address: SRI International, 201 Washington Rd, Princeton, NJ 08540.


{}

\end{document}